# Development of IoT Sensor and Cloud-based Server for Cloud-based Bridge Long-term Monitoring

*Junyoung Park, Junsik Shin & Jongwoong Park*
Chung-Ang University, Republic of Korea

**ABSTRACT:** As social infrastructures rapidly age, it is crucial to create a digital SOC (Social Overhead Capital) maintenance system for preventive maintenance. Using IoT sensors installed on the structures, abnormal signals produced by the structures may be rapidly identified and the best judgments can be made. In this study, a multimetric IoT sensor was integrated into a digital SOC monitoring system for long-term monitoring, use in a cloud computing server for automated and powerful data analysis, and for creating databases to perform: (1) multimetric sensing, (2) long-term operation, and (3) LTE-based direct communication. The created sensor included five strain sensing axes and three acceleration axes, and it also had an event-driven power management system that only turned on the sensors when the vibration reached a certain threshold or the timer started. Long-term operation might be made possible by the power management system and an extra solar panel charging. Real-time data transmission to the server from the sensors was accomplished using low-power LTE CAT.1 connection, which does not need an additional gateway device. Additionally, a displacement fusion technique was created on the cloud server to get reference-free structural displacement for ambient structural assessment after receiving multi-variable data from the sensor. A steel railroad and concrete girder bridge were used to experimentally evaluate the suggested digital SOC system.

**KEYWORDS:** Smart IoT Sensor, Long-term Monitoring, LTE, Displacement Estimation, Cloud Computing, Equivalent Neutral Axis, Structural Health Monitoring

## 1. Introduction

Bridges are structures that play a pivotal role in economic development as a basis for various production activities. In Korea, regular monitoring is carried out for special bridges that use the cables connected to the pylons to lift the bridge decks, but most general bridges are verified their safety through precise safety diagnosis every 6 months to 2 years. Precise safety diagnosis is to inspect facilities in accordance with the 'Special Act on Safety Management of Facilities', and in the case of bridges, dynamic characteristics are checked through impulse response. However, as in the case of corrosion of Cheongdam 1 Bridge in 2020 and the tendon damage of Jeongleungcheon in 2016, the bridge deteriorates over time and unexpected damage may occur with consuming a lot of time and money. In addition, due to the increasing number of aging bridges and a decrease in construction manpower, there is a great difficulty in the inspection speed, and long-term measures are required (Messore et al., 2020).

To overcome the difficulties of the manpower-based inspection method, studies have been conducted to continuously monitor the bridge using IoT sensors (Lee et al., 2017). Monitoring using IoT sensors has the advantages of relatively low sensor cost and ease of installation. However, the existing IoT sensor uses a communication method with a short communication range, such as zigbee or bluetooth, or it takes a lot of time to transmit data because the communication speed is slow. In addition, the existing IoT sensor is a vibration-based sensor, and it was difficult to measure displacement, which is a major indicator of the dynamic characteristics of a bridge, as it only collects acceleration and natural frequencies (Zarafshan et al., 2012).

This study introduces the long-term monitoring of bridges through IoT sensors and cloud servers. Section 2 presents the design of self-developed IoT sensor, JANET. JANET is equipped with a high-performance micro controller unit (MCU) that can quickly process various calculations, an Analog Digital Converter (ADC) for high-precision measurement, and an Event Driven System (EDS) circuit to minimize power consumption. JANET is a stand-alone sensor that wirelessly transmits data to the server using LTE Cat 1 communication. Section 3 describes the cloud server that processes the collected data and the cloud built-in algorithm for calculating displacement from the collected acceleration and strain. Section 4 illustrates the contents of attaching such a sensor system to Cheongdam 1 Bridge, comparing its performance with LVDT, and performing long-term monitoring. Also, accuracy of the sensor's displacement calculation and various analysis results are performed from data measured over 3 months.

## 2. System Design: JANET

This section describes about a sensor named JANET, which was developed for long-term bridge monitoring. JANET includes an ADC that can simultaneously measure multi-channel data at high resolution, a MCU that can process many operations quickly, and LTE modem that can wirelessly transmit measured data to the server. In addition, the EDS circuit and solar charging system are equipped so that it can perform semi-permanently without battery replacement.

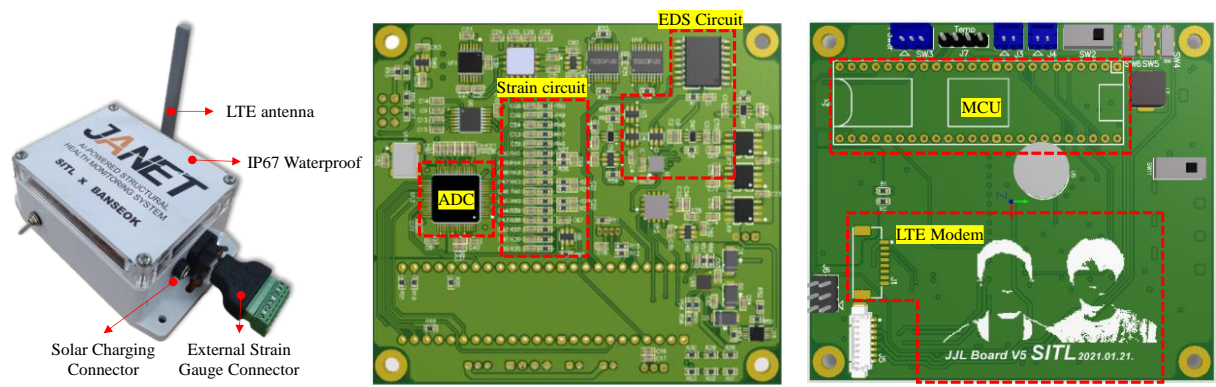

Fig. 1: (a) Developed JANET and (b) its PCB

### 2.1 Multi-metric Sensing Circuit

#### 2.1.1 MCU

Table 1: Comparison of Micro Controller Unit

|  | Teensy 4.1 | Arduino Due | Arduino Mega | Adafruit Metro M4 | Sparkfun ESP32 Thing |
|---|---|---|---|---|---|
| Core | ARM Cortex-M7 | ARM Cortex-M3 | ATmega2560 | ARM Cortex M4 | Dual core Tensilica LX6 |
| Clock Speed | 600 MHz | 84 MHz | 16 MHz | 120 MHz | 240 MHz |
| RAM | 9 MB | 96 KB | 8 KB | 192 KB | 520 KB |
| SDIO | Micro SD Socket | - | - | - | - |
| Serial Ports | 8 | 4 | 4 | 1 | 3 |
| SPI Ports | 3 | 1 | 1 | 1 | 3 |
| I2C Ports | 3 | - | - | 1 | 2 |

In the case of Imote2, SHM-A, which were developed for wireless bridge monitoring in the past, the sensors could not be developed further due to production suspension by manufacturers (Park et al., 2011). To prevent these problems in advance, JANET adopted Arduino-based MCU, an open-source hardware, where the circuit diagram, printed circuit board, and hardware description language are all open to the public. There were two more considerations for MCU selection. These were fast clock speed for wireless data transmission and adequate size of RAM for multi-channel data acquisition. Among the various Arduino-based MCUs, Teensy 4.1 comes to a consensus about the considerations, featuring high specifications (see table 1). Teensy 4.1 provides extensive memory availability of 7936kB and 1024kB respectively in flash and RAM with additional 8MB PSRAM, enabling extra data storage (Teensy® 4.1, n.d.). Also, it is built with 32-bit ARM Cortex-M7 core with Arduino IDE support, running at 600 MHz clock speed, bringing enough processing speed.

### 2.1.2 High Sensitivity Multi-Channel Data Acquisition

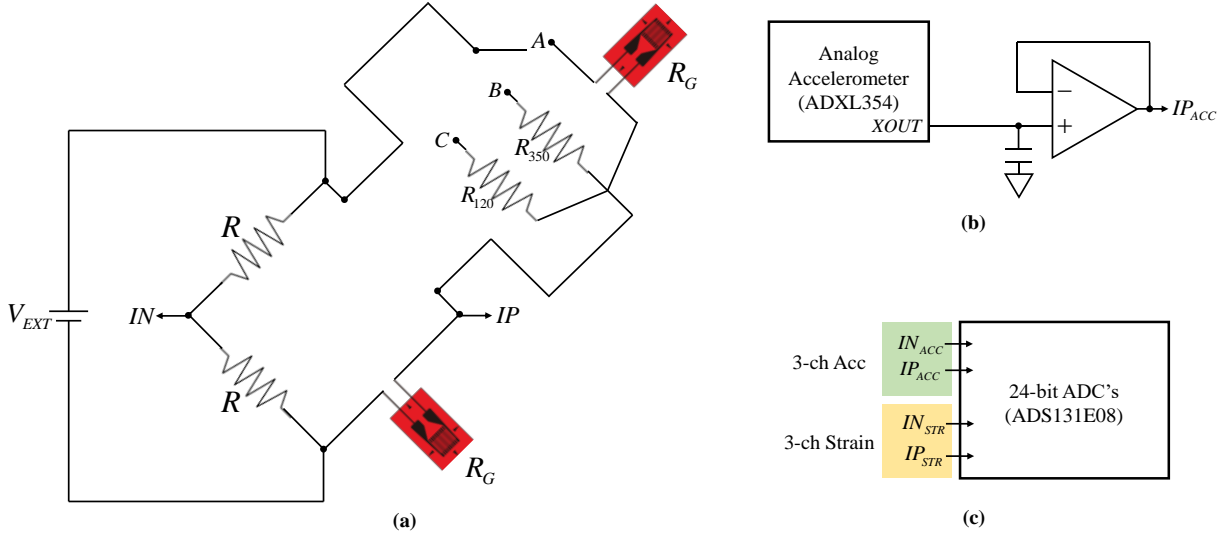

Fig. 2: (a) Strain sensing circuit, (b) Acceleration sensing circuit and (c) ADC differential mode

Figure 2 shows configurations of developed analog sensing circuits. JANET acquires strain and acceleration data with high fidelity owing to deployment of wheatstone bridge circuit and 8 channel 24-bit analog to digital converter (ADC) (ADS131E08, n.d.). Although the wheatstone bridge circuit provides precise analog signal measurements, the fact that strain gauges are sensitive to temperature causes signal fluctuation and drift. Unlike the concrete bridge where drift is negligible as it has less temperature impact, drift in steel bridge is relatively more affected by temperature (Khan et al., 2021). To this end, a hardware switch was built into JANET so that wheatstone circuit could be set as a half-bridge when attaching a steel bridge for temperature compensation while single bridge (120 Ω/350 Ω) is utilized for concrete bridge. The dummy gauge used for the half-bridge is mounted 90-degrees to the sensitive axis of an active strain gauge to represent only the change due to temperature. Since both gauges are equally affected by temperature, this method helps to efficiently offset the drift. After through the wheatstone circuit, $IN_{STR}$, $IP_{STR}$ are inputted to ADC as a differential mode input. Note that $IN_{STR}$ is fixed with half of the excitation voltage ($V_{EXT}$)

The ADXL354 is a three-axis microelectromechanical system (MEMS) accelerometer with ultra-low noise density of 20 μg/√Hz. Also, the device only consumes 150 μA in measurement mode. It provides a low noise of 0.15 mg with a bandwidth of 50 Hz under the ±2 g ranges. Similar with differential mode in strain sensing circuit, $IN_{ACC}$ is fixed by shorted with opampin and opampout of ADC pins while $IP_{ACC}$ is the varying voltage from ADXL354, an analog accelerometer. Accelerometer herein measures X-axis, Y-axis, and Z-axis simultaneously so three channels are allocated to the ADC.

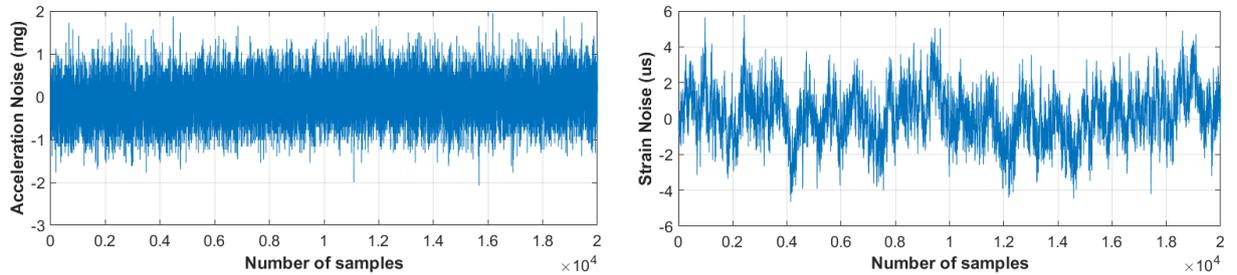

Fig. 3: Noise level of JANET: (a) acceleration and (b) strain

The noise level of JANET was validated via a laboratory-scale experiment. The tests for acceleration and strain were conducted for 20 seconds data with a sampling rate of 1000Hz. Acceleration was measured in vibration-free site and strain gauge were attached at the top of a 1.5-high cantilever beam. The result of Root Mean Square Error (RMSE) related to acceleration and strain were 0.45 mg and 1.52 μs, respectively. These confirmed that JANET can measure acceleration and strain over 0.45 mg and 1.52 μs.

## 2.2 Event-driven System for power management

A wireless sensor that is always on would quickly deplete its battery. However, once the sensor attached to any infrastructures, it is difficult to replace its battery due to structure's poor accessibility causes additional cost for ladder truck or traffic control. Therefore, the solution proposed herein is to use a trigger mechanism, so called EDS. EDS has been widely reported in the literature for its efficient power management strategy (Sarwar et al., 2020). In this sensor, EDS is based on two types of triggering, vibration-based triggering, and timer-based triggering. Figure 4 shows the control logic design of the EDS circuit.

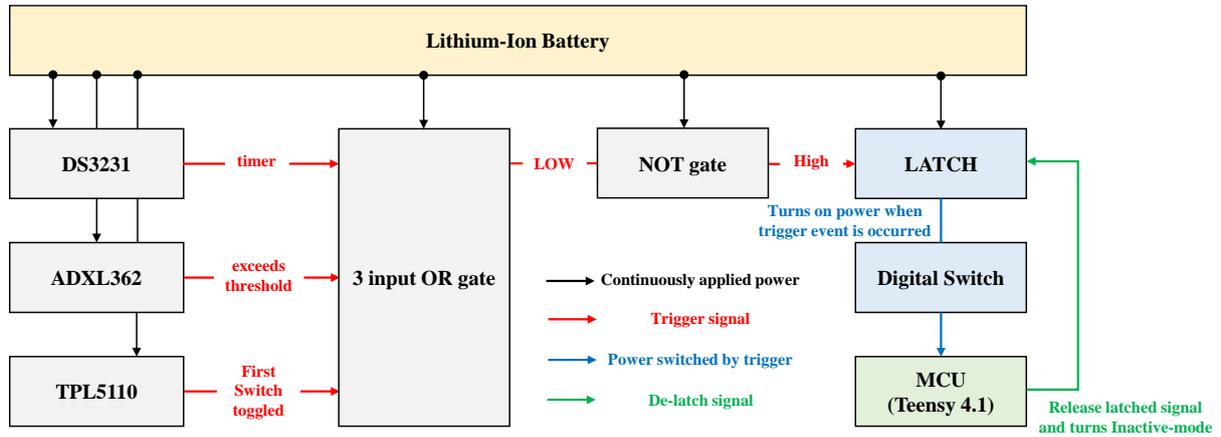

Fig. 4: Control logic diagram of EDS

ADXL362 is an ultra-low power 3-axis MEMS accelerometer. When an event occurs and exceeds pre-defined threshold, the interrupt pin of ADXL362 generates a trigger signal for user-defined time. DS3231 is a Real Time Clock (RTC) module that helps to initialize the smart sensor to reach time synchronization with real world time. Similar with ADXL362, the interrupt pin of RTC generates a trigger signal when user-defined time is met. To generate the combined triggering signals from ADXL362 and RTC, a universal gate SN74AUP1G58 from Texas Instruments (TI) was used. Since ADXL362 is designed active high, whereas DS3231 is designed active low, the universal gate output is adjusted to generate low signal (active low) whenever one of them is triggered. This low signal is then inverted to a high signal using a not gate and delivered to latch.

The latch has two inputs and one output, latch enable (LE), D and Q, respectively. When LE is high, Q follows D. When LE is low then Q retains the previous state of D. Initially, LE is pulled up to pass the trigger signal and Q is low because there is no trigger signal (D is low). Once the trigger signal is generated, D goes high, which also makes Q high. This activates the digital switch and turns on the MCU. However, the trigger signal is reset in a few seconds for next event and if the signal goes low when LE is high, the MCU is shut down. To prevent this, MCU makes the LE pin low at the very beginning to maintain a high signal, until process ends. When the process is completed, the MCU turns the LE pin back to high and goes to inactive mode.

## 2.3 Software Framework

An integrated software framework for JANET was developed under Arduino-based open-source drivers. Before entering software implementation, JANET maintains a low-power state in inactive mode until a trigger event occurs. Once triggered by timer or vibration, JANET in active mode measures 3-channel acceleration and 3-channel strain at 1000Hz for 30 seconds. After measurement, high-frequency noise is removed through FIR filtering and down sampled by 10 times. After that, JANET tries to connect to the cloud server up to 10 times for wireless data transmission through LTE. When connection succeeds, MCU generates to JavaScript Object Notation (JSON) formed data packet and transmits it to the server. By AT commands, acknowledgement (ACK) response returns whenever the packet is sent successfully. At this time, if data transmission fails, JANET repeats infinitely until data is sent to the server, thereby minimizing data loss. And when all data is transmitted, the sensor returns to inactive mode. Figure 5 describes whole process of the software.

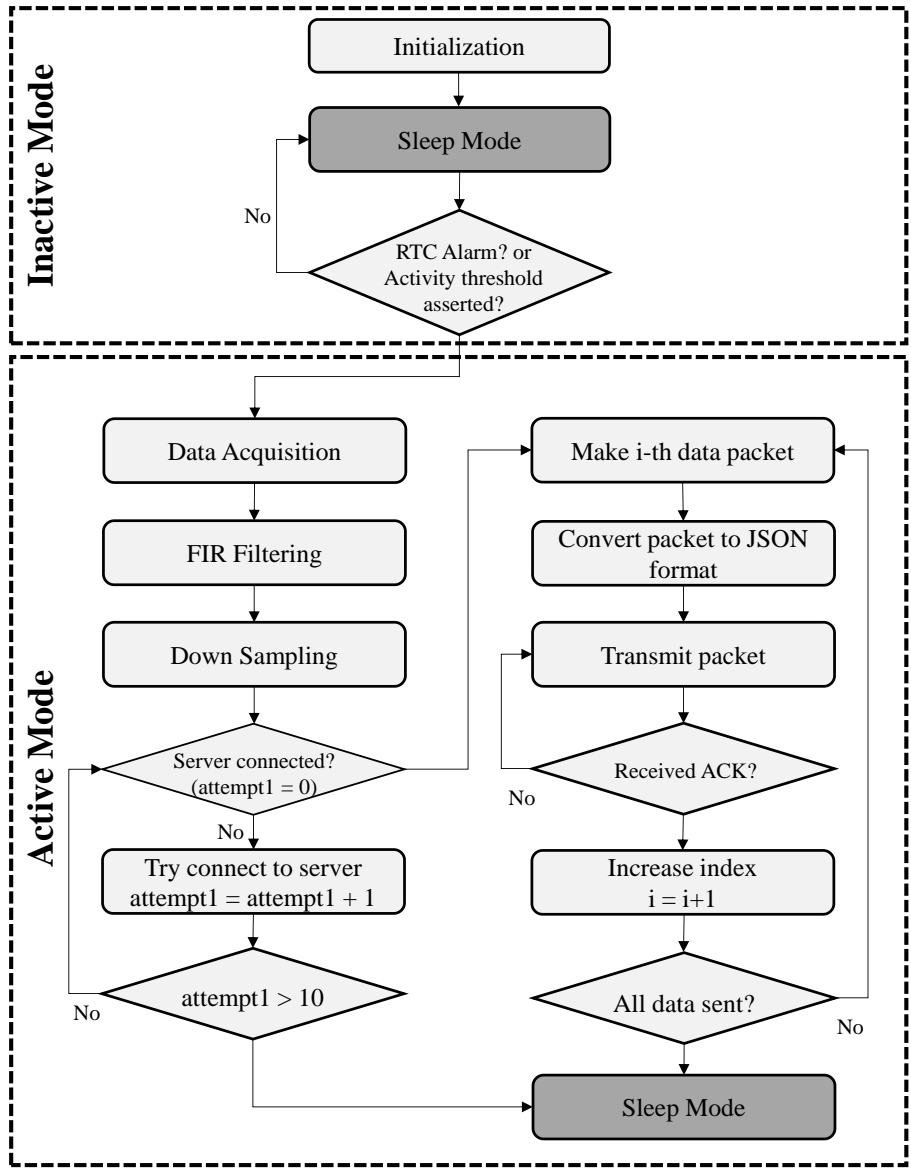

Fig. 5: Developed software flow for JANET

## 2.4 Wireless Data Acquisition through TCP/IP using LTE Cat 1

To make JANET as a stand-alone sensor, which can work without application of additional gateway or router, LTE Cat 1, a wireless communication standard designed for IoT applications, was used. LTE Cat 1 modem adopted herein is P500S manufactured by Woori-Net in Republic of Korea for its convenience and simplicity. P500S is supplied 5V and provides two UARTs, one USB serial, and one external GNSS Passive antenna. Moreover, AT Commands that can control the modem with prescribed rules are provided by the manufacturers.

After data acquisition is ended, JANET transmits measured data to cloud server with the P500S. The AT command of the P500S has embedded TCP/IP communication composed of IP, which is a packet communication type internet protocol, and TCP, which is a transmission control protocol. To satisfy the input parameter type of AT Commands, the data should be packaged to JSON objects. To convert the data into JSON format, ArduinoJSON, an open-source software supporting efficient JSON serialization was used.

If whole 6-channel data measured for 30 seconds at the rate of 100Hz is made into a single JSON packet, packet generation can be simply done. However, transmission may fail due to its large size. If a packet is created at every data point (3000 times in total), on the other hand, the time required for the data transmission becomes too long, which causes excessive power consumption. After a lot of trials and errors, it is adopted to bundle one JSON packet for every 8 data points. Figure 6 shows explicit process using TCP/IP commands.

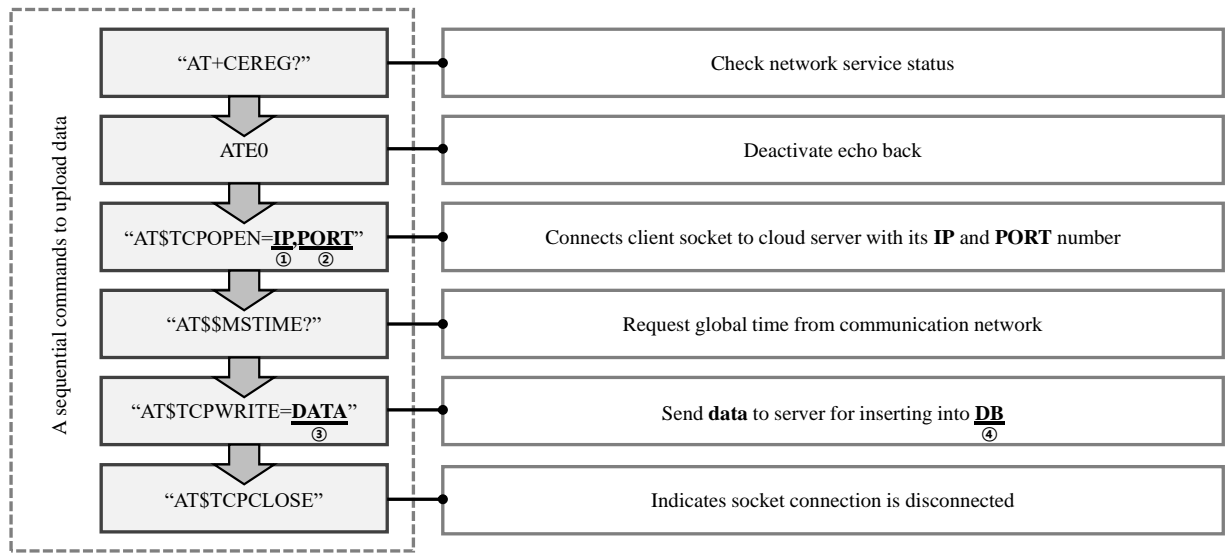

Fig. 6: Block diagram of TCP/IP commands for uploading data

- **IP:** Public IPv4 address of Amazon EC2 instance designed for cloud computing (e.g., "127.0.0.1")
- **PORT:** Number assigned to logical construct that identifies a specific process (e.g., "3306" for MySQL service)
- **DATA:** Sequence of characters converted hexadecimal value, including JSON formatted measured data (e.g., 434155 for "CAU" converted 0x43, 0x41, 0x55)
- **DB:** Table name of MySQL database a key of JSON object (e.g., "CHEONGDAM1_info")

When an AT Command is sent, an ACK response is returned each time, indicating whether the command was successfully executed. For minimizing data loss, it was developed to retry infinitely when the packet transmission is failed by checking the ACK response for each packet send.

## 2.5 Power consumption of JANET

JANET performs data acquisition, processing, and transmission to the cloud server based on the trigger-signal. In active mode the current consumption averages 300 mA. The current consumptions were measured by USB power meter that provides voltage readings down to 0.01V and current to 0.001A. Whereas in inactive mode, current of 4.191 mA is consumed (see Table 2).

Table 2: Current consumption in Inactive mode

| Device | Current Consumption |
| --- | --- |
| ZS-042 (DS3231) | 200 μA |
| SN74AUP1G58 (2) | 1.8 μA |
| SN74LVC1G373DCKR | 10 μA |
| ADXL362 | 1.8 μA |
| TPL5110_timer | 35 nA |
| TPS22919 | 8 μA |
| MCP1725-3302E/SN (3) | 360 μA |
| TPS22860 (3) | 6 nA |
| LT3652 | 2.5 mA |
| LTC2990 | 1.1 mA |
| SN74AUP1G58 (2) | 1.8 µA |
| TPS22919 | 8 µA |
| ADG734 (2) | 40 nA |
| Total | 4.191 mA |

A total of three 18650 lithium-ion rechargeable batteries were deployed to have the battery capacity up to 10,350 mAh. The battery life of JANET that stands for time (day) is computed as equation 1.1:

$$Battery = \frac{Capacity(mAh)}{I_{AVG}} \times 0.8 \qquad (1.1)$$

where $Battery$ represents the battery life in hour, $I_{AVG}$ represents the average current consumption and 0.8 is adopted as a compensation factor for environmental and material conditions affecting battery's life (Khan et al., 2021). $I_{AVG}$ can be calculated using equation 1.2.

$$I_{AVG} = I_{in-Active}(1 - nT_s) + I_{Active}(nT_s) \qquad (1.2)$$

Here, $T_s$ is sensing time (230 s). Considering an operation of six times a day, JANET has an $I_{AVG}$ equal to 214.02 mA and $Battery$ of 38.7 h (1.62 days). Besides, solar panel that installed with JANET in the field recharges lithium-ion batteries automatically at a current rate of 550 mA when the sun is up. Therefore, if the solar panel is charged for more than 9 hours a day, the supplied current is greater than the current consumed by the sensor, which means JANET can operate semi-permanently without replacing the battery.

## 3. Data Pipeline for Cloud-based Processing

### 3.1 Overview of the Cloud Server

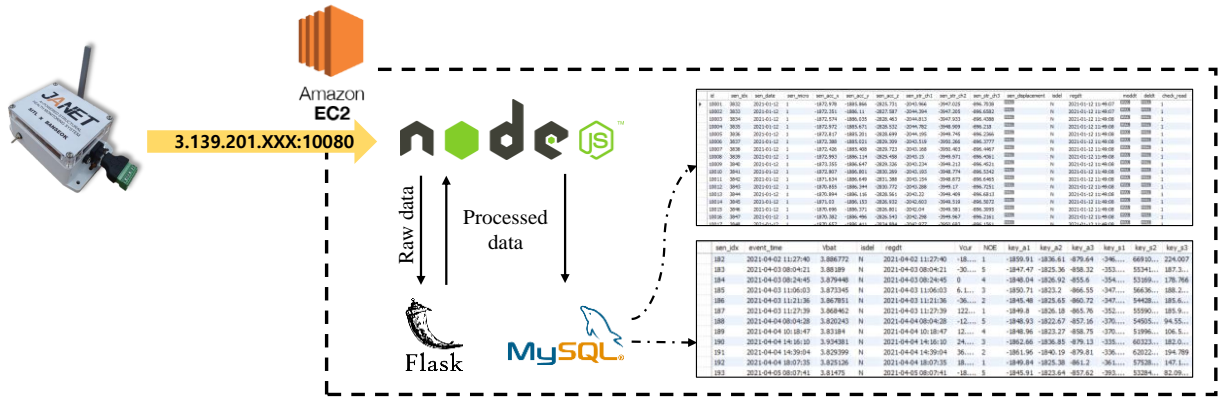

Fig. 7: Cloud server system diagram

Amazon EC2 is adopted as the cloud server of JANET. It provides a virtual computing environment, so called instance, to build a server and configure a network without any hardware for high performance computing. Since the server can opened by designating a country/region, it has strong scalability to collect sensor data anywhere in the world with a proper LTE modem for the country. Instance serves various elements that make up the server, such as CPU, memory, storage, and network capacity, and use a security group to specify the protocol, port, and source IP range, enabling customized server development. In this study, according to the TCP/IP command of the LTE modem introduced in Figure 6, a static IP for the server was issued using the Elastic IP Address (EIP) service. Furthermore, by using port forwarding to designate a port for DB connection, measured data for each JANET can be properly stored. Figure 7 shows the detailed workflow performed on the server. The backend server designed with Node.js is always on, and when data is received, it calls python-based Flask to perform various cloud computing operations. This analyzed data is stored in a MySQL database (DB). The DB consists of a DB that stores measured sensor data, analyzed data, and a DB that stores the state of the sensor such as battery voltage, solar charging current, etc. (Park et al., 2021)

### 3.2 Embedded Displacement Estimation Algorithm

Displacement, which is a key indicator of bridge dynamic characteristics, is not easy to measure as fixed point is required to install a displacement gauge and it is difficult to obtain such a fixed point for river bridges or long-span bridges (Won et al., 2021). On behalf of using direct displacement-measuring system, displacement is indirectly estimated using the multimetric technique that fuses the acceleration and strain proposed by Park et al. (See Eq. 1.3).

$$u_{strain} = \Phi q = \Phi \Psi \varepsilon = \alpha \Phi \Phi'' \varepsilon \tag{1.3}$$

where $u_{strain}$ is the strain-based displacement, $\Phi$ and $\Psi$ are the mode shape matrix of the strain and displacement, respectively.

Then, a strain-based displacement with the same shape as the actual displacement was obtained. To scale the strain-based displacement to the actual displacement unit, it is required to be multiplied by $\alpha$, a scaling factor. This scaling factor is calculated by the equation given below, which is obtained by dividing the power spectral density (PSD) of the strain-based displacement and the PSD of the acceleration-based displacement.

$$\alpha = \sqrt{\frac{S_{strain,xi}^{disp}(f_n)}{S_{acc,xi}^{disp}(f_n)}} \tag{1.4}$$

where $S_{strain,xi}^{disp}$ is the PSD of the strain-based displacement, $S_{acc,xi}^{disp}$ is the PSD of the acceleration-based displacement and $f_n$ is the dominant natural frequency.

The $\alpha$ did not regarded simply as a scaling factor but as a physical quantity for the equivalent neutral axis of the bridge (Park et al., 2019). Also, the value could perform an indicator of the load-bearing performance of the bridge. In this paper as well, $\alpha$ is obtained in the process of estimating the displacement, so the load-bearing performance of the bridge was compared using this value. This is introduced in Section 4.2.

## 4. Implementation of the Long-term Monitoring to Road Bridge

### 4.1 Bridge Description

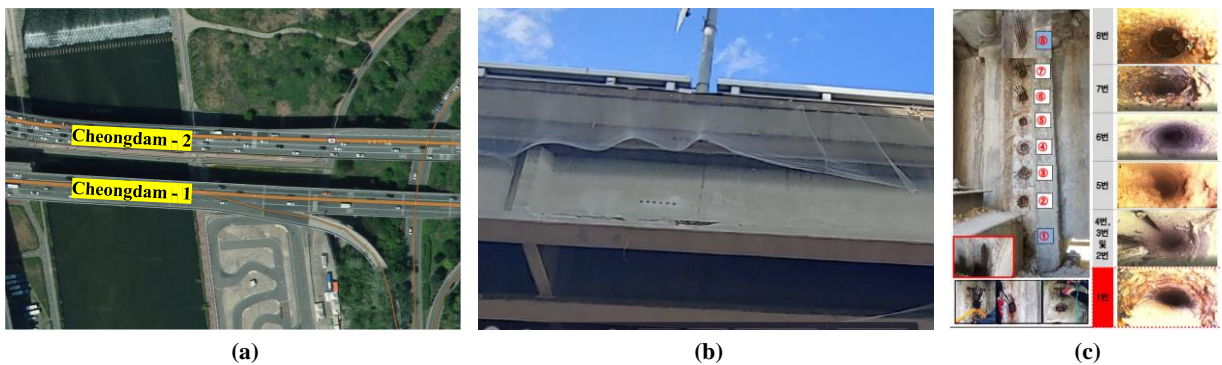

(a)          (b)          (c)

Fig. 8: (a) Birds eye view of Cheongdam 1 bridge, (b) Damaged bridge surface and (c) Corroded tendon

Cheongdam 1 Bridge in Seoul is a PSC girder bridge, and in September 2020, cracks due to tendon corrosion were discovered and repairs were made. Tendon corrosion has been frequently reported since 1990 in Europe, where PSC girder bridges were applied at least 30 years earlier, but in Korea, it was the first time since the tendon corrosion of Jeongneungcheon Viaduct in 2016. For this reason, Cheongdam 1 Bridge was a bridge with the meaning of announcing the beginning of the aging of PSC bridges in Korea. JANET was installed in Cheongdam 1 bridge from July 21, 2021, and performed short-term and long-term monitoring, respectively.

Short-term monitoring was performed on the day of installation to verify the displacement measurement performance of the sensor. LVDT and one JANET were installed at the same time to compare the response of the same location for passing vehicles, and through this, the displacement measurement accuracy of the sensor was verified. Long-term monitoring was performed without LVDT, with only two JANETs attached, and two performance verifications were conducted. The first is long-term operation performance verification. EDS circuit, solar charging and debugging algorithm for fault-tolerance have been verified to work well and the sensor doesn't stop during operation. The second is a comparison of the load carrying capacity of the girders. Since the two JANETs were installed on the S2-G3 girder where damage was not reported and the S3-G1 girder where damage was reported, respectively, by comparing the data collected from the two sensors, it can be predicted whether repair and reinforcement were properly performed. During long-term monitoring, the threshold value for the vibration-based trigger was set to 200 mg, and the timer-based trigger was set to occur once a day at 8 am. In addition, measurement was performed for 30 seconds at the time of one trigger.

### 4.2 Comparison Displacements from LVDT with JANET

Figure 9 shows the estimated displacement calculated from the acceleration and strain and the reference displacement from LVDT. The location where LVDT was installed is same with where JANET attached to compare results. Three local maximums of displacement were noted for difference calculation. The three maximum displacement values measured from the LVDT were 2.08 mm, 1.01 mm, and 1.26 mm, respectively, and the values calculated through JANET were 2.18 mm, 1.09 mm, and 1.31 mm, indicating that displacement estimation is available within an error range of 0.1 mm.

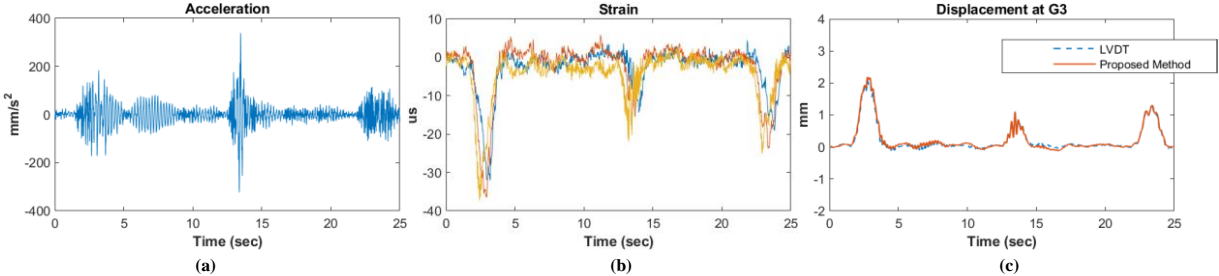

Fig. 9: (a) z-axis acceleration, (b) three channels strain and (c) comparison of structural displacements

Table 1: Current consumption in Inactive mode

| LVDT | JANET | Differences |
| --- | --- | --- |
| 2.08 mm | 2.18 mm | 0.1 mm |
| 1.01 mm | 1.09 mm | 0.08 mm |
| 1.26 mm | 1.31 mm | 0.05 mm |

### 4.3 Long-term Data Analysis

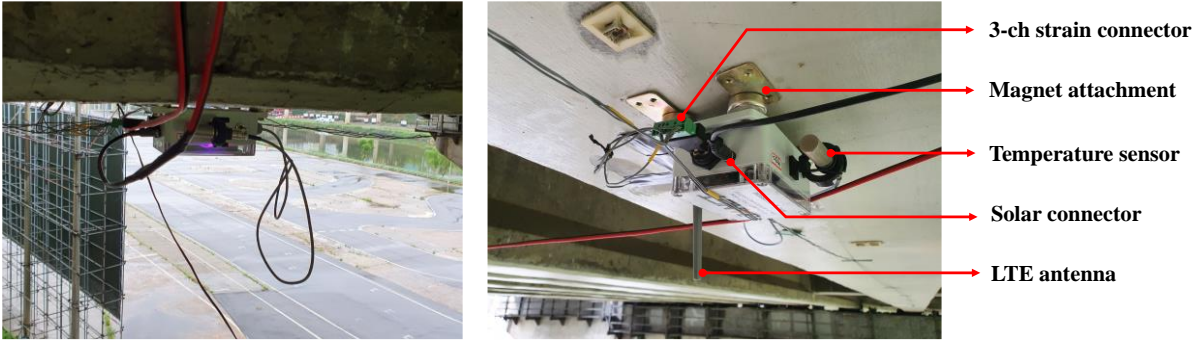

Fig. 10: Sensor deployment

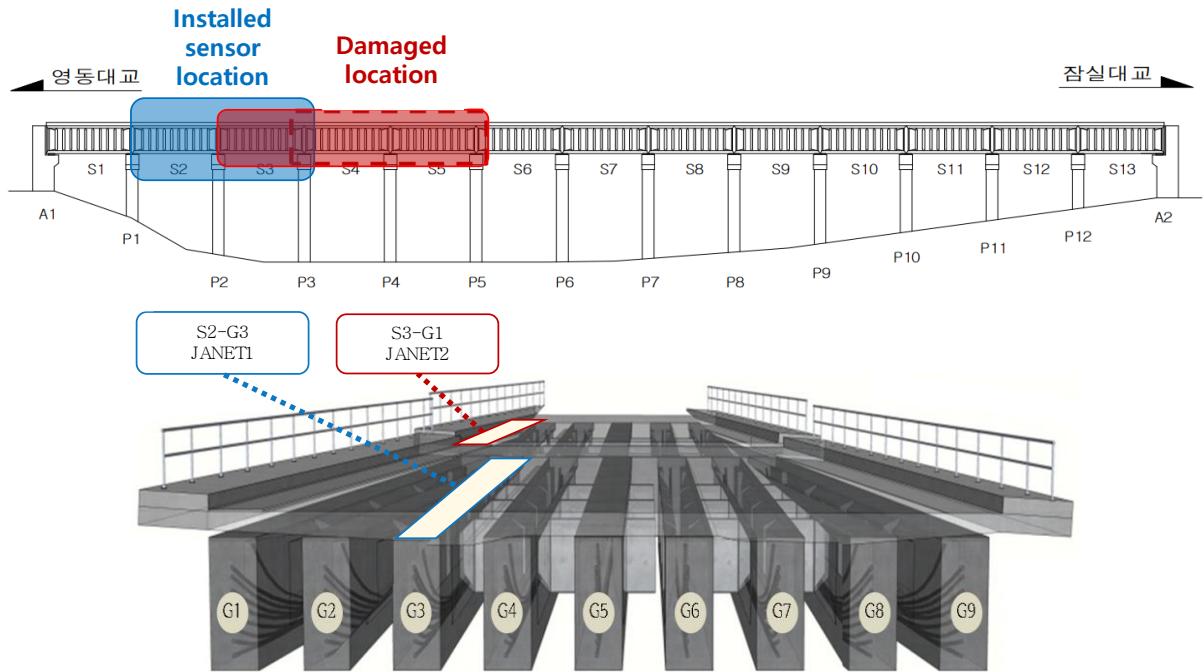

Fig. 11: Experimental setup

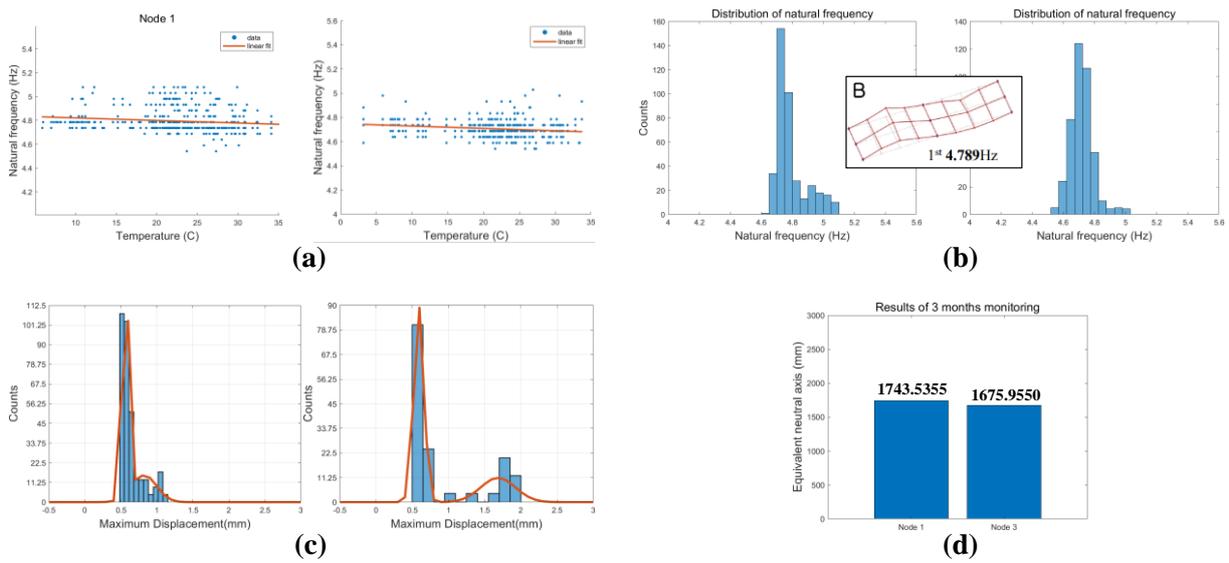

Fig. 12: (a) Relationships between temperature and natural frequency (b) histogram of natural frequencies, (c) multivariate gaussian distribution of maximum displacements and (d) equivalent neutral axis

Figure 12 (a) shows the relationship between the measured temperature and natural frequency for 3 months was analyzed. The temperature was obtained from a thermometer attached to the side of the sensor, and the natural frequency was obtained from the z-direction acceleration. Each trend line is presented in equations (1.5) and (1.6). The natural frequency is slightly decreased as temperature increased. It indicates that the elastic modulus of concrete and the rigidity of the structure decrease as the temperature increases by (Lee et al., 2003), and consequently the natural frequency decreases.

$$y = -0.0021x + 4.8420 \tag{1.5}$$

$$y = -0.002x + 4.7488 \tag{1.6}$$

Figure 12 (b) shows the natural frequencies extracted from the z-direction accelerations of the two sensors, respectively. Although it changes little by little through the analysis results, it can be confirmed that 4.78 Hz and 4.78 Hz are dominant, and these results are almost consistent with the 4.789 Hz known through as the natural frequency of the same bridge.

Figure 12 (c) shows the maximum displacement distribution calculated through cloud server. The data of 0.5mm or less are scarce, because most of the data acquisition is event-based. So, all vibration data below the set threshold were not collected. The distribution acquired from the two sensors is not a simple normal distribution, but a multivariate distribution composed of multivariate Gaussian distribution components. Through this, it can be expected that the traffic vehicle that caused the trigger is composed of two types of heavy vehicles and freight vehicles.

The scaling factors, which indicates the equivalent neutral axis obtained from the data for each sensor are 1743.54 mm and 1675.96 mm, respectively. Sensors attached to different girders, one of which is the girder for which girder damage has been reported, both of which have similar values. That is, it indicates that the repair and reinforcement of the bridge was properly performed.

## 5. Conclusions

In this paper JANET, an IoT sensor developed for cloud-based long-term monitoring of bridges, was introduced and a cloud server for data collection and high-performance computational processing. The developed IoT sensor adopts Teensy 4.1 as MCU to process high-performance calculation and performs high-precision measurement of 3-axis acceleration and 3-channel strain through 8-channel ADC. In addition, to minimize power consumption, a novel EDS circuit that is triggered based on acceleration and timer is embedded. The measured data is transmitted through TCP/IP communication to the server through LTE Cat 1, a low-power IoT communication. It is equipped with a low-power circuit and solar charging so that it can be operated for a long time without replacing the battery. The server was built with Amazon ec2 instance, an Amazon cloud service. The instance is always on to enable 24-hour cloud processing, and when data access is attempted to the node.js backend server, it automatically calculates displacement, natural frequency, and equivalent neutral axis from the measured acceleration and strain through the Python-based flask algorithm. Afterwards, the measurement data and processing data are automatically linked to the MySQL-based DB.

JANET was installed and applied for 3 months from July 2021 at Cheongdam 1 Bridge, which was repaired after damage due to tendon corrosion reported in 2020. On the day of installation for long-term monitoring, Janet and LVDT were installed together to compare displacement, and it was verified that displacement was precisely measured with an error of less than 0.1 mm. Afterwards, two sensors were installed, one each on the damaged girder and the normal girder. Through the setup, various analysis using long-term monitoring were performed.

As a future work, the power circuit of the sensor will be further optimized, and additional analysis algorithms such as BIM will be built in the cloud. In addition, by applying sensors to new bridges, new analyzes will be derived from long-term data of multiple bridges.

## 6. Acknowledgements

This work was supported by the Korea Agency for Infrastructure Technology Advancement (KAIA) grant funded by the Ministry of Land, Infrastructure and Transport (Grant 22CTAP-C164014-02).

## 7. REFERENCES (STYLE HEADING 2)

ADS131E08. (n.d.). Retrieved September 25, 2022, from https://www.ti.com/document-viewer/ADS131E08/datasheet/abstract#SBAS561376

Khan, S., Won, J., Shin, J., Park, J., Park, J. W., Kim, S. E., ... & Kim, D. J. (2021). SSVM: An Ultra-Low-Power Strain Sensing and Visualization Module for Long-Term Structural Health Monitoring. Sensors, 21(6), 2211.

Lee, J. L., Tyan, Y. Y., Wen, M. H., & Wu, Y. W. (2017, May). Development of an IoT-based bridge safety monitoring system. In 2017 international conference on applied system innovation (ICASI) (pp. 84-86). IEEE.

Lee, S. C. (2003). Prediction of concrete strength using artificial neural networks. Engineering structures, 25(7),

849-857.

Messore, M. M., Capacci, L., & Biondini, F. (2020). Life-cycle cost-based risk assessment of aging bridge networks. Structure and Infrastructure Engineering, 17(4), 515-533.

Park, J. H., Kim, J. T., & Lee, S. Y. (2011, February 28). Performance Evaluation of Imote2-Platformed Wireless Smart Sensor Node for Health Monitoring of Harbor Structures. Journal of Korean Society of Coastal and Ocean Engineers, 23(1), 26–33.

Park, J. W., Moon, D. S., Sim, S. H., & Spencer Jr, B. F. (2019). Equivalent neutral axis for structural condition assessment using multi-sensor fusion. Engineering Structures, 197, 109350.

Park, J. Y., Shin, J. S., Won, J. B., Park, J. W., & Park, M. Y. (2021). Development of Low-Power IoT Sensor and Cloud-Based Data Fusion Displacement Estimation Method for Ambient Bridge Monitoring. Journal of the Computational Structural Engineering Institute of Korea, 34(5), 301-308.

Sarwar, M. Z., Saleem, M. R., Park, J. W., Moon, D. S., & Kim, D. J. (2020). Multimetric event-driven system for long-term wireless sensor operation for SHM applications. IEEE Sensors Journal, 20(10), 5350-5359.

Teensy® 4.1. (n.d.). Retrieved September 25, 2022, from https://www.pjrc.com/store/teensy41.html

Won, J., Park, J. W., Park, J., Shin, J., & Park, M. (2021). Development of a Reference-Free Indirect Bridge Displacement Sensing System. Sensors, 21(16), 5647.

Zarafshan, A., Iranmanesh, A., & Ansari, F. (2012, November). Vibration-Based Method and Sensor for Monitoring of Bridge Scour. Journal of Bridge Engineering, 17(6), 829–838.